\documentclass[aps,preprint,showpacs,floats,epsf,epsfig,nofootinbib,12pt]{revtex4}
\usepackage{graphicx}% Include figure files
\usepackage{epsfig}
\usepackage[dvips]{color}

\def\beq{\begin{eqnarray}}
\def\eeq{\end{eqnarray}}
\def\non{\nonumber}

\def\la{\langle}
\def\ra{\rangle}
\def\Mb{M_{\mathcal{B}_1}}
\def\Mc{M_{\mathcal{B}_2}}

\begin{document}

\title{ Revisiting the transition $\Xi^{+}_{cc}\to\Xi^{(')+}_c$ to understand the data from LHCb}

\vspace{1cm}

\author{ Hong-Wei Ke$^{1}$\footnote{khw020056@tju.edu.cn} and
        Xue-Qian Li$^2$\footnote{lixq@nankai.edu.cn}
   }

\affiliation{  $^{1}$ School of Science, Tianjin University, Tianjin 300072, China
\\
  $^{2}$ School of Physics, Nankai University, Tianjin 300071, China }

\vspace{12cm}

\begin{abstract}
The LHCb collaboration newly measured the decay rate of doubly charmed baryon $\Xi^{++}_{cc}\to\Xi^{'+}\pi^+$
and a ratio of its branching fraction with respect to that of the decay $\Xi^{++}_{cc}\to\Xi^{+}\pi^+$ is reported as
$1.41\pm 0.17\pm 0.10$. This result conflicts with the
theoretical predictions made by several groups. In our previous work, following the prescription given in early literature where
the $us$ diquark in $\Xi^{+}_{c}$ is assumed to be a scalar
whereas in $\Xi^{'+}_{c}$ is a vector i.e. the spin-flavor structure of $\Xi^{+}_{c}$ is $[us]_0 c$ and that of  $\Xi^{'+}_{c}$ is $[us]_1 c$, we studied the case of $\Xi^{++}_{cc}\to\Xi^{(')+}$ with the light front quark model. Numerically we obtained $\Gamma(\Xi^{++}_{cc}\to\Xi^{'+}\pi^+)/\Gamma(\Xi^{++}_{cc}\to\Xi^{+}\pi^+)=0.56\pm0.18$ which is about  half of the data.
While abandoning the presupposition, we suppose the spin-flavor structure of $us$ in $\Xi^{+}_{c}$  may be a mixture of scalar
and vector, namely the spin-flavor function of $\Xi^{+}_{c}$  could be ${\rm cos}\theta\, [us]_{0}[c]+{\rm sin}\theta\, [us]_{1}[c]$. An alternative combination
$-{\rm sin}\theta\,[us]_{0}[c]+{\rm cos}\theta\, [us]_{1}[c]$ would correspond to $\Xi^{'+}_{c}$.  Introducing the mixing mechanism the ratio $\Gamma(\Xi^{++}_{cc}\to\Xi^{'+}\pi^+)/\Gamma(\Xi^{++}_{cc}\to\Xi^{+}\pi^+)$ depends on the mixing
angle $\theta$.
With the mixing scenario, the theoretical prediction on the ratio between the transition rate of $\Xi^{+}_{cc}\to\Xi^{'+}_c$ and that of $\Xi^{+}_{cc}\to\Xi^{+}_c$ can coincide with the data as long as  $\theta=16.27^\circ\pm2.30^\circ$ or $85.54^\circ\pm2.30^\circ$ is set. Definitely, more precise measurements on other decay portals of $\Xi^{+}_{cc}$ are badly needed for testing the mixing mechanism
and further determining the mixing angle.

\end{abstract}

\maketitle

\section{Introduction}

In 2017, the LHCb collaboration observed the doubly charmed baryon
$\Xi^{++}_{cc}$\cite{Aaij:2017ueg} in the four-body final
state $\Lambda_cK^{-}\pi^+\pi^+$ \cite{Aaij:2017ueg} and the $\Xi^{+}\pi^+$
portals \cite{Aaij:2018gfl} successively. That baryon was expected
for a long time by physicists of high energy physics because of its significance.
The quark model predicted existence of baryons with two or three heavy quarks but they had evaded from experimental observation
for long.
With the great progress of detecting facilities and techniques, recently the LHCb collaboration observed the doubly charmed baryon via
a decay portal $\Xi^{++}_{cc}\to\Xi^{'+}\pi^+$
with a branching fraction relative to that of the decay $\Xi^{++}_{cc}\to\Xi^{+}\pi^+$ being $1.41\pm 0.17\pm 0.10$ \cite{LHCb:2022rpd}.

On the theoretical aspect, some approaches have been applied to study the weak decay $\Xi^{++}_{cc}\to\Xi^{(')+}$.
In Refs. \cite{Wang:2017mqp,Shi:2019hbf,Sharma:2017txj,Gerasimov:2019jwp,Cheng:2020wmk,Gutsche:2018msz,Gutsche:2019iac,Ivanov:2020xmw,Gutsche:2017hux}
the predicted ratio $\Gamma(\Xi^{++}_{cc}\to\Xi^{'+}\pi^+)/\Gamma(\Xi^{++}_{cc}\to\Xi^{+}\pi^+)$ was not in keeping with the data. In our earlier work \cite{Ke:2021rxd} the transition $\Xi^{++}_{cc}\to\Xi^{(')+}$ was explored within the
light front quark model \cite{Jaus,Ji:1992yf,Cheng:1996if,Cheng:2003sm,Lu:2007sg,
Li:2010bb,Ke:2009ed,Ke:2010,Choi:2007se,
Ke:2009mn,Ke:2011fj,Ke:2011mu,Ke:2011jf,Chua:2018lfa,Yu:2017zst,Ke:2007tg,
Wei:2009np,Ke:2012wa,Ke:2017eqo,Ke:2019smy,pentaquark1,pentaquark2,Tawfiq:1998nk,Chang:2018zjq,Geng:2022xpn}, where the three-quark
picture of baryon was adopted. In that approach one needs to determine the vertex functions for the baryons by means of
their inner structures. For $\Xi^{++}_{cc}$ a naive and
reasonable conjecture suggests that the two $c$
quarks compose a physical subsystem  (or a diquark)  which serves as a color source for the light quark
\cite{Falk:1993gb,Chang:2007xa}. The relative orbital angular momentum between the two
$c$ quarks is 0, i.e. the $cc$ pair is in an $S$ wave, and because it is in a color-anti-triplet $\bar 3$,
the spin of the $cc$ pair must be 1 due to the
symmetry requirement. In the works about single charmed-baryons, usually the two light
quarks are supposed to reside in a subsystem as the light
diquark \cite{Ebert:2006rp,Korner:1992wi}. In those literatures, a presupposition
Ref.\cite{Ebert:2006rp} suggests that the $us$ diquark in $\Xi^{+}_{c}$ is  a scalar
whereas a vector in $\Xi^{'+}_{c}$.

In the transition $\Xi^{++}_{cc}\to\Xi^{(')+}$ one $c$ quark in the initial state would decay into an $s$ quark via weak interaction
while the other $c$ quark and $u$ quark are spectators in the process but the $cu$ pair is not a diquark (or a physical subsystem) in the initial state, neither in the final state.
To utilize the spectator scenario, the quark structure of $[cc]_1u$ ($[us]_0 c$ or $[us]_1 c$) should be mathematically rearranged into a sum of $\sum_i c[cu]_i$ ($\sum_i s[cu]_i$)
where the sum runs over all possible spin projections via a Racah
transformation. It is found that the spectator $cu$ is independent of the quark ($c$ or $s$) which is involved in the transition process.
Thus the transition process can be divided into two steps: First, the
physical structure  $[cc]_1u$ is rearranged into $[cu]_i c$ by a Racah transformation and then  the single $c$ quark decays into $s$ by emitting a gauge boson while
the subsystem of $[cu]_i$ remains unchanged; Second, the $[cu]_i s$
structure in the final state is re-ordered into the $[us]_0 c$ or $[us]_1 c$  structure through another Racah transformation.
In our ealier work on the
transition $\Xi^{++}_{cc}\to\Xi^{(')+}$ \cite{Ke:2021rxd} the three quarks are treated as
individual subjects i.e. possess their own momenta, and we obtained the branching ratio $\Gamma(\Xi^{++}_{cc}\to\Xi^{'+}\pi^+)/\Gamma(\Xi^{++}_{cc}\to\Xi^{+}\pi^+)$ as $0.56\pm0.18$
which is almost half of the data.

It is noted that the earlier calculation in \cite{Ke:2021rxd} was based on the presupposition that the $us$ diquark in $\Xi^{+}_{c}$ is  a scalar
whereas that in $\Xi^{'+}_{c}$ is a vector \cite{Ebert:2006rp}.
As is well known the flavor symmetry is broken so either the $us$ diquark in $\Xi^{+}_{c}$ or
in $\Xi^{'+}_{c}$ could be a mixture of a scalar and a vector. In this new scenario a mixing angle $\theta$ ($0< \theta<\pi$) should be introduced for the mixing
of flavor-spin wave functions i.e. $\Xi^{+}_{c}={\rm cos}\theta\, [us]_{0}[c]+{\rm sin}\theta\, [us]_{1}[c]$ and $\Xi^{'+}_{c}=-{\rm sin}\theta\,[us]_{0}[c]+{\rm cos}\theta\, [us]_{1}[c]$ where the subscript 0 or 1 represents the total spin of the $us$ subsystem. When $\theta$ is set to
0, i.e.  $\Xi^{+}_{c}= [us]_{0}[c]$ and $\Xi^{'+}_{c}= [us]_{1}[c]$, the structures of $\Xi^{+}_{c}$ and $\Xi^{'+}_{c}$ restore the original setting supposed by the authors of Ref.\cite{Ebert:2006rp}.
In other words the transition matrix element $\mathcal{A}(\Xi^{++}_{cc}\to\Xi^{+}_{c})$ and $\mathcal{A}(\Xi^{++}_{cc}\to\Xi^{'+}_{c})$ carried out in Ref.\cite{Ke:2021rxd} just correspond to the processes $\mathcal{A}([cc]_{1}[u]\to[us]_{0}[c])$
and $\mathcal{A}([cc]_{1}[u]\to[us]_{1}[c])$.
Now as long as  $\theta$ is not equal to  0, the process for $\mathcal{A}(\Xi^{++}_{cc}\to\Xi^{+}_{c})$ should be replaced by ${\rm cos}\theta\, \mathcal{A}([cc]_{1}[u]\to[us]_{0}[c])+ {\rm sin}\theta\, \mathcal{A}([cc]_{1}[u]\to[us]_{1}[c]$), while for
$\mathcal{A}(\Xi^{++}_{cc}\to\Xi^{'+}_{c})$ the process is  $-{\rm sin}\theta\, \mathcal{A}([cc]_{1}[u]\to[us]_{0}[c])+ {\rm cos}\theta\, \mathcal{A}([cc]_{1}[u]\to[us]_{1}[c]$). The simple extension means existence of new sub-transition matrix elements for
$\Xi^{++}_{cc}\to\Xi^{+}_{c}$ and $\Xi^{++}_{cc}\to\Xi^{'+}_{c}$.  The mixture scenario changes the predicted rate from the old picture,
obviously the newly obtained theoretical estimate on the rates  depend on $\theta$.

This paper is
organized as follows: after the introduction, in section II we revisit the transition matrix element for $\Xi^{+}_{cc}\rightarrow \Xi^{(')+}_{c}$
in the light-front quark model. Our numerical results for $\Xi^{+}_{cc}\rightarrow
\Xi^{(')+}_{c}$ are presented in section
III. The section IV is devoted to our conclusion and discussions.

\section{$\Xi^{+}_{cc}\rightarrow \Xi^{(')+}_{c}$ in the light-front quark model}

\subsection{The structures of $\Xi^{+}_{cc}$, $\Xi^{+}_{c}$ and $\Xi^{'+}_{c}$}

The spectator scenario may greatly
alleviates the theoretical difficulties for calculating the hadronic transition matrix elements.
However the diquarks (physical subsystems) $cc$ and  $us$ in $\Xi^{+}_{cc}$ and $\Xi^{(')}_{c}$ are not spectators, which means the spectator approximation cannot be directly applied.
In fact the $c$ and $u$ quarks
which do not undergo a transition in the process, i.e. the combination of $cu$
is approximately regarded as a spectator (an effective subsystem).

As a
three-body system, the total spin of a baryon can be realized by different
constructing schemes and the Racah transformations can relate one to others.
By the aforementioned rearrangement of quark flavors the physical
states $[cc]_1 u$  and $[us]_{0}c$ (or $[us]_{1}c$) are written into sums over the
effective forms $c[cu]_i$ and $s[cu]_i$, respectively.
The detailed transformations are\cite{Wang:2017mqp}
\begin{eqnarray}
&&{[c^1c^2]}_{1}[u]=\frac{\sqrt{2}}{2}(-\frac{\sqrt{3}}{2}[c^2][c^1u]_{0}+
\frac{1}{2}[c^2][c^1u]_{1}\nonumber
\\&&\,\,\,\,\,\,\,\,\,\,\,\,\,\,\,\,\,\,\,\,\,\,\,\,\,\,\,\,\,\,\,\,\,\,\,\,
\,\,\,\,-\frac{\sqrt{3}}{2}[c^1][c^2u]_{0 }+\frac{1}{2}[c^1][c^2u]_{1}),\\
&&[us]_{0}[c]=-\frac{1}{2}[s][cu]_{0}+\frac{\sqrt{3}}{2}[s][cu]_{1},\\
&&[us]_{1}[c]=\frac{\sqrt{3}}{2}[s][cu]_{0}+\frac{1}{2}[s][cu]_{1}.
\end{eqnarray}
In Ref. \cite{Ebert:2006rp} $\Xi^{+}_{c}\equiv{[c^1c^2]}_{1}[u]$, $\Xi^{+}_{c}\equiv[us]_{0}[c]$ and $\Xi^{'+}_{c}\equiv[us]_{1}[c]$,
instead, in this work we suppose $\Xi^{+}_{c}$ and $\Xi^{'+}_{c}$ to be mixtures of $[us]_{0}[c]$ and $[us]_{1}[c]$ i.e.
$\Xi^{+}_{c}={\rm cos}\theta\, [us]_{0}[c]+{\rm sin}\theta\, [us]_{1}[c]$ and $\Xi^{'+}_{c}=-{\rm sin}\theta\,
[us]_{0}[c]+{\rm cos}\theta\, [us]_{1}[c]$ where $\theta$ is the mixing angle (restricted in the first and second quadrants with
$0< \theta<\pi$).

\subsection{ the  form factors of $\Xi^{+}_{cc}\to\Xi^+_c$ and $\Xi^{+}_{cc}\to\Xi^{'+}_c$ in LFQM}

\begin{figure}
\begin{center}
%\begin{tabular}{ccc}
\scalebox{0.6}{\includegraphics{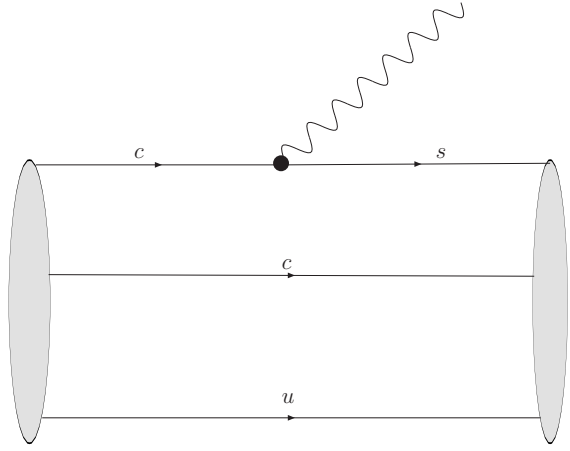}}
%\end{tabular}
\end{center}
\caption{The Feynman diagrams for $\Xi^{+}_{cc}\to\Xi^{(')+}_{c}$ transitions,
where $\bullet$ denotes $V-A$ current vertex.}\label{t1}
\end{figure}

The leading Feynman diagram responsible for the weak decay
$\Xi^{+}_{cc}\to\Xi_c^{(')+}$ is shown in Fig. \ref{t1}. Following the
procedures given in
Refs. \cite{pentaquark1,pentaquark2,Ke:2007tg,Ke:2012wa} the
transition matrix element can be computed with the vertex
functions of $\mid \Xi^{+}_{cc}(P,S,S_z) \ra$ and $\mid
\Xi_{c}^{(')+}(P',S',S'_z) \ra$. The $cu$ subsystem stands as a
spectator, i.e. its spin configuration does not change during the
transition.

The form factors for the weak transition $\Xi^{+}_{cc}\rightarrow
\Xi^{+}_{c}$  are defined in the standard way as
\begin{eqnarray}\label{s1}
&& \la \Xi^{+}_{c}(P',S',S_z') \mid \bar{s}\gamma_{\mu}
 (1-\gamma_{5})c \mid \Xi^{+}_{cc}(P,S,S_z) \ra  \non \\
 &=& \bar{u}_{\Xi^{+}_{c}}(P',S'_z) \left[ \gamma_{\mu} f_{1}(q^{2})
 +i\sigma_{\mu \nu} \frac{ q^{\nu}}{M_{\Xi^{+}_{cc}}}f_{2}(q^{2})
 +\frac{q_{\mu}}{M_{\Xi^{+}_{cc}}} f_{3}(q^{2})
 \right] u_{\Xi^{+}_{cc}}(P,S_z) \nonumber \\
 &&-\bar u_{\Xi^{+}_{c}}(P',S'_z)\left[\gamma_{\mu} g_{1}(q^{2})
  +i\sigma_{\mu \nu} \frac{ q^{\nu}}{M_{\Xi^{+}_{c}}}g_{2}(q^{2})+
  \frac{q_{\mu}}{M_{\Xi^{+}_{cc}}}g_{3}(q^{2})
 \right]\gamma_{5} u_{\Xi^{+}_{cc}}(P,S_z).
\end{eqnarray}
where  $q \equiv P-P'$. In terms of the spin-flavor structures of $\Xi^{+}_{cc}$ and $\Xi^{+}_{c}$
the matrix element $\la \Xi^{+}_{c}(P',S',S_z') \mid \bar{s}\gamma_{\mu}
 (1-\gamma_{5})c \mid \Xi^{+}_{cc}(P,S,S_z) \ra$ can be written as
 $${\rm cos}\theta\la [s][cu]_{0}
\mid \bar{s}\gamma_{\mu}
 (1-\gamma_{5})c \mid [c][cu]_{0} \ra+{\rm sin}\theta\la [s][cu]_{1} \mid
\bar{s}\gamma_{\mu}
 (1-\gamma_{5})c \mid [c][cu]_{1} \ra.$$   For the transition matrix elements $\la [s][cu]_{0}
\mid \bar{s}\gamma_{\mu}
 (1-\gamma_{5})c \mid [c][cu]_{0} \ra$ and $\la [s][cu]_{1} \mid
\bar{s}\gamma_{\mu}
 (1-\gamma_{5})c \mid [c][cu]_{1} \ra$ the form factors are denoted to $f_i^s$, $g_i^s$  and $f_i^v$ , $g_i^v$, so we have
\begin{eqnarray}\label{relation}
f_1=(\frac{\sqrt{6}}{4}{\rm cos}\theta-\frac{3\sqrt{2}}{4}{\rm sin}\theta)f^s_1+(\frac{\sqrt{6}}{4}{\rm cos}\theta+\frac{\sqrt{2}}{4}{\rm sin}\theta)f^v_1,\nonumber\\
g_1=(\frac{\sqrt{6}}{4}{\rm cos}\theta-\frac{3\sqrt{2}}{4}{\rm sin}\theta)g^s_1+(\frac{\sqrt{6}}{4}{\rm cos}\theta+\frac{\sqrt{2}}{4}{\rm sin}\theta)g^v_1,\nonumber\\
f_2=(\frac{\sqrt{6}}{4}{\rm cos}\theta-\frac{3\sqrt{2}}{4}{\rm sin}\theta)f^s_2+(\frac{\sqrt{6}}{4}{\rm cos}\theta+\frac{\sqrt{2}}{4}{\rm sin}\theta)f^v_2,\nonumber\\
g_2=(\frac{\sqrt{6}}{4}{\rm cos}\theta-\frac{3\sqrt{2}}{4}{\rm sin}\theta)g^s_2+(\frac{\sqrt{6}}{4}{\rm cos}\theta+\frac{\sqrt{2}}{4}{\rm sin}\theta)g^v_2,
\end{eqnarray}\label{relation}
and $f_i^s$, $g_i^s$, $f_i^v$,
$g_i^v$ can be found in our earlier paper\cite{Ke:2019smy} .

For the transition $\la \Xi^{'+}_{c}(P',S',S_z') \mid
\bar{s}\gamma_{\mu}
 (1-\gamma_{5})c \mid \Xi^{+}_{cc}(P,S,S_z) \ra$
the form factors are also defined as done in Eq. (\ref{s1}). Here we
just add a symbol `` $'$ " on $f_1$, $f_2$, $g_1$ and $g_2$  to
distinguish the quantities for $\Xi^{+}_{cc}\to\Xi_c$ and those for
$\Xi^{+}_{cc}\to \Xi'_c$. They are
\begin{eqnarray}\label{relation}
f'_1=(-\frac{\sqrt{6}}{4}{\rm sin}\theta-\frac{3\sqrt{2}}{4}{\rm cos}\theta)f^s_1+(-\frac{\sqrt{6}}{4}{\rm sin}\theta+\frac{\sqrt{2}}{4}{\rm cos}\theta)f^v_1,\nonumber\\
g'_1=(-\frac{\sqrt{6}}{4}{\rm sin}\theta-\frac{3\sqrt{2}}{4}{\rm cos}\theta)g^s_1+(-\frac{\sqrt{6}}{4}{\rm sin}\theta+\frac{\sqrt{2}}{4}{\rm cos}\theta)g^v_1,\nonumber\\
f'_2=(-\frac{\sqrt{6}}{4}{\rm sin}\theta-\frac{3\sqrt{2}}{4}{\rm cos}\theta)f^s_2+(-\frac{\sqrt{6}}{4}{\rm sin}\theta+\frac{\sqrt{2}}{4}{\rm cos}\theta)f^v_2,\nonumber\\
g'_2=(-\frac{\sqrt{6}}{4}{\rm sin}\theta-\frac{3\sqrt{2}}{4}{\rm cos}\theta)g^s_2+(-\frac{\sqrt{6}}{4}{\rm sin}\theta+\frac{\sqrt{2}}{4}{\rm cos}\theta)g^v_2.
\end{eqnarray}\label{relation}

\section{Numerical Results}

\subsection{The form factors for $\Xi^{+}_{cc}\to \Xi^+_c$ and $\Xi^{+}_{cc}\to \Xi^{'+}_c$  }

In Ref. \cite{Ke:2021rxd} we used a polynomial to parameterize these form factors  $f^s_i$, $g^s_i$, $f^v_i$ and $g^v_i$ ($i=1,2$),
 \begin{eqnarray}\label{s146}
 F(q^2)=F(0)\left[1+a\left(\frac{q^2}{M_{\Xi^{+}_{cc}}^2}\right)
  +b\left(\frac{q^2}{M_{\Xi^{+}_{cc}}^2}\right)^2+c\left(\frac{q^2}{M_{\Xi^{+}_{cc}}^2}\right)^3\right].
 \end{eqnarray}
The fitted values of $a,~b,~c$ and $F(0)$ in the form factors
are presented in Table \ref{Tab:t2}.
With the form factors, we re-estimate the decay rates of
semi-leptonic and no-leptonic decays with the new scenario for the diquark structures.

\begin{table}
\caption{The form factors given in
 polynomial form.}\label{Tab:t2}
\begin{ruledtabular}
\begin{tabular}{ccccc}
  $F$    &  $F(0)$ &  $a$  &  $b$ & $c$\\\hline
  $f^s_1$  &   0.586    & 1.57    &1.59 & 0.704   \\
$f^s_2$  &   -0.484     &2.06  &2.42 &1.17   \\
  $g^s_1$  &      0.420    &    0.983& 0.692& 0.258  \\
  $g^s_2$  &      0.228  &    1.90  &  2.07 &0.960  \\
  $f^v_1$  &   0.610     &  2.04    &2.27 & 1.06   \\
$f^v_2$  &   0.463     &  2.14    & 2.49 & 1.19  \\
  $g^v_1$  &      -0.140    &    0.422 & 0.0931 & 0.00632 \\
  $g^v_2$  &      0.0673   &    0.925  &  0.245 & -0.0862
\end{tabular}
\end{ruledtabular}
\end{table}

\begin{figure}
\begin{center}
%\begin{tabular}{ccc}
\scalebox{0.8}{\includegraphics{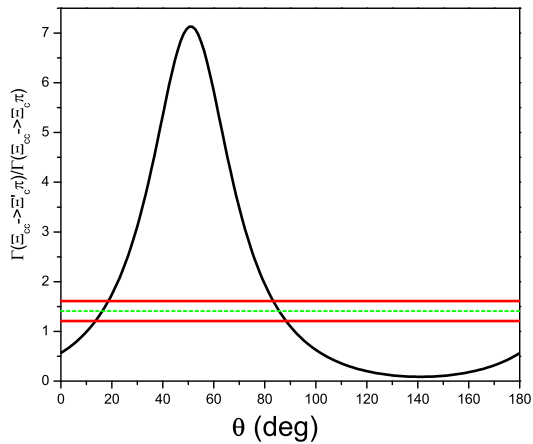}}
%\end{tabular}
\end{center}
\caption{The dependence of  $\Gamma(\Xi^{+}_{cc}\to\Xi^{+}_{c}\pi)/\Gamma(\Xi^{+}_{cc}\to\Xi^{'+}_{c}\pi)$ on $\theta$.}\label{fdg0}
\end{figure}

\subsection{Non-leptonic decays of ${\Xi^{+}_{cc}}\to\Xi^{+}_{c}+ M$ and ${\Xi^{+}_{cc}}\to\Xi^{'+}_{c}+ M$}

The transition matrix element
of the non-leptonic decay is complicated due to involving non-perturbative physics.
We did the calculation in Ref. \cite{Ke:2021rxd} by employing the
factorization assumption,
\begin{eqnarray}\label{s0}
&& \la \Xi^{(')+}_{c}(P',S_z')M \mid \mathcal{H} \mid \Xi^{+}_{cc}(P,S_z) \ra  \nonumber \\
 &=&\frac{G_FV_{cs}V^*_{qq'}}{\sqrt{2}}f_M\la
\Xi^{(')+}_{c}(P',S_z') \mid \bar{s} \gamma^{\mu} (1-\gamma_{5})c
\mid \Xi^{+}_{cc}(P,S_z) \ra,
\end{eqnarray}
where $f_M$ is the decay constant of meson $M$.
Besides the decay rate, the up-down asymmetry
parameter $\alpha$ (its definition can be found in Appendix) is also an experimentally observable quantity which has obvious significance for understanding
the governing mechanism (including information about the non-perturbative effects). In our following tables we explicitly
offer the theoretically estimated values for $\alpha$ corresponding to different mixing angles.

Using the Eq. (\ref{s0}) we show the dependence of the ratio $\frac{\Gamma(\Xi^{+}_{cc}\to\Xi^{+}_{c}\pi)}{\Gamma(\Xi^{+}_{cc}\to\Xi^{'+}_{c}\pi)}$ on $\theta$
which is depicted in Fig. \ref{fdg0} (the horizontal band centered with the dotted line corresponds to the range allowed by the experimental error tolerance).
With the data $1.41\pm0.17\pm0.10$ we fix $\theta$ to be $16.27^\circ\pm2.30^\circ$ or $85.54^\circ\pm2.30^\circ$. The mixing angle deviates from $0^\circ$, which might manifest the scale of the flavor SU(3) symmetry breaking for the concerned processes.

The mixing mechanism can change the predictions on the non-leptonic decays significantly. 
We list the estimated decay rates and up-down asymmetries of those processes with the three mixing angles in
Tabs. \ref{Tab:theta0}, \ref{Tab:theta1}, \ref{Tab:theta2}. Comparing the results shown in the three tables one can find:

1. the orders of magnitude for the channels are unchanged with or without the mixing;

2. $\Xi^{+}_{cc}\to\Xi^{(')+}_{c} \pi$  and $\Xi^{+}_{cc}\to\Xi^{(')+}_{c}
\rho$ are the main two-body decay channels for $\Xi^{+}_{cc}$;

3. the relative sizes between $\Gamma(\Xi^{+}_{cc}\to\Xi^{+}_{c} M)$ and $\Gamma(\Xi^{+}_{cc}\to\Xi^{'+}_{c} M)$ are varied for the different mixing angles.

%In order to determine which mixing angle is correct  more decay channels of $\Xi^{+}_{cc}$ should be measured.

\begin{table}
\caption{The Widths (in unit $10^{10}{\rm s}^{-1}$)
and up-down asymmetry of non-leptonic decays
$\Xi^{+}_{cc}\to\Xi^{(')+}_{c} M$ in \cite{Ke:2021rxd} ($\theta=0^\circ$).}\label{Tab:theta0}
\begin{ruledtabular}
\begin{tabular}{ccc|ccc}
 mode& width & up-down asymmetry &mode &width  &up-down asymmetry\\\hline
  $\Xi^{+}_{cc}\to\Xi^{+}_{c} \pi$ & 13.6$\pm$1.8&-0.441$\pm$0.009&$\Xi^{+}_{cc}\to\Xi^{'+}_{c} \pi$  &7.68$\pm$0.92 &-0.982$\pm$0.005 \\\hline
  $\Xi^{+}_{cc}\to\Xi^{+}_{c} \rho$ &11.0$\pm$1.5&-0.429$\pm$0.016&
  $\Xi^{+}_{cc}\to\Xi^{'+}_{c} \rho$&13.9$\pm$1.2&-0.111$\pm$0.034\\\hline
  $\Xi^{+}_{cc}\to\Xi^{+}_{c} K$   &1.03$\pm$0.14&-0.402$\pm$0.008&$\Xi^{+}_{cc}\to\Xi^{'+}_{c}
  K$&0.492$\pm$0.059&-0.998$\pm$0.002
                            \\\hline
  $\Xi^{+}_{cc}\to\Xi^{+}_{c} K^{*}$ &
  0.414$\pm$0.055&-0.422$\pm$0.021&
  $\Xi^{+}_{cc}\to\Xi^{'+}_{c} K^{*}$&0.623$\pm$0.052&-0.014$\pm$0.030
\end{tabular}
\end{ruledtabular}
\end{table}

\begin{table}
\caption{The Widths (in unit $10^{10}{\rm s}^{-1}$)
and up-down asymmetry of non-leptonic decays
$\Xi^{+}_{cc}\to\Xi^{(')+}_{c} M$ with $\theta=16.27^\circ\pm2.30^\circ$.}\label{Tab:theta1}
\begin{ruledtabular}
\begin{tabular}{ccc|ccc}
 mode& width & up-down asymmetry &mode &width  &up-down asymmetry\\\hline
  $\Xi^{+}_{cc}\to\Xi^{+}_{c} \pi$ & 8.37$\pm$0.69&-0.087$\pm$0.070&$\Xi^{+}_{cc}\to\Xi^{'+}_{c} \pi$  &11.8$\pm$0.5 &-0.991$\pm$0.006 \\\hline
  $\Xi^{+}_{cc}\to\Xi^{+}_{c} \rho$ &5.59$\pm$0.56&-0.167$\pm$0.079&$\Xi^{+}_{cc}\to\Xi^{'+}_{c} \rho$&17.6$\pm$0.3&-0.228$\pm$0.014\\\hline
  $\Xi^{+}_{cc}\to\Xi^{+}_{c} K$   &0.642$\pm$0.052&-0.081$\pm$0.063&$\Xi^{+}_{cc}\to\Xi^{'+}_{c} K$&0.789$\pm$0.041&-0.967$\pm$0.010\\\hline
  $\Xi^{+}_{cc}\to\Xi^{+}_{c} K^{*}$ &0.187$\pm$0.021&-0.211$\pm$0.084&$\Xi^{+}_{cc}\to\Xi^{'+}_{c} K^{*}$&0.756$\pm$0.011&-0.107$\pm0.011$
\end{tabular}
\end{ruledtabular}
\end{table}

\begin{table}
\caption{The Widths (in unit $10^{10}{\rm s}^{-1}$)
and up-down asymmetry of non-leptonic decays
$\Xi^{+}_{cc}\to\Xi^{(')+}_{c} M$ with $\theta=85.54^\circ\pm2.30^\circ$.}\label{Tab:theta2}
\begin{ruledtabular}
\begin{tabular}{ccc|ccc}
 mode& width & up-down asymmetry &mode &width  &up-down asymmetry\\\hline
  $\Xi^{+}_{cc}\to\Xi^{+}_{c} \pi$ & 8.36$\pm$0.69&-0.952$\pm$0.023&$\Xi^{+}_{cc}\to\Xi^{'+}_{c} \pi$  &11.8$\pm$0.6 &-0.507$\pm$0.032 \\\hline
  $\Xi^{+}_{cc}\to\Xi^{+}_{c} \rho$ &16.8$\pm$0.9&-0.106$\pm$0.023&$\Xi^{+}_{cc}\to\Xi^{'+}_{c} \rho$&8.23$\pm$0.67&-0.456$\pm$0.004\\\hline
  $\Xi^{+}_{cc}\to\Xi^{+}_{c} K$   &0.554$\pm$0.049&-0.977$\pm$0.039&$\Xi^{+}_{cc}\to\Xi^{'+}_{c} K$&0.869$\pm$0.038&-0.455$\pm$0.029\\\hline
  $\Xi^{+}_{cc}\to\Xi^{+}_{c} K^{*}$ &0.834$\pm$0.042&-0.005$\pm$0.013&$\Xi^{+}_{cc}\to\Xi^{'+}_{c} K^{*}$&0.269$\pm$0.027&-0.419$\pm$0.008
\end{tabular}
\end{ruledtabular}
\end{table}

\subsection{Semi-leptonic decays of $\Xi^{+}_{cc} \to
\Xi^{+}_{c} +l\bar{\nu}_l$ and $\Xi^{+}_{cc} \to \Xi^{'+}_{c} +l\bar{\nu}_l$}

\begin{figure}[hhh]
\begin{center}
\scalebox{0.7}{\includegraphics{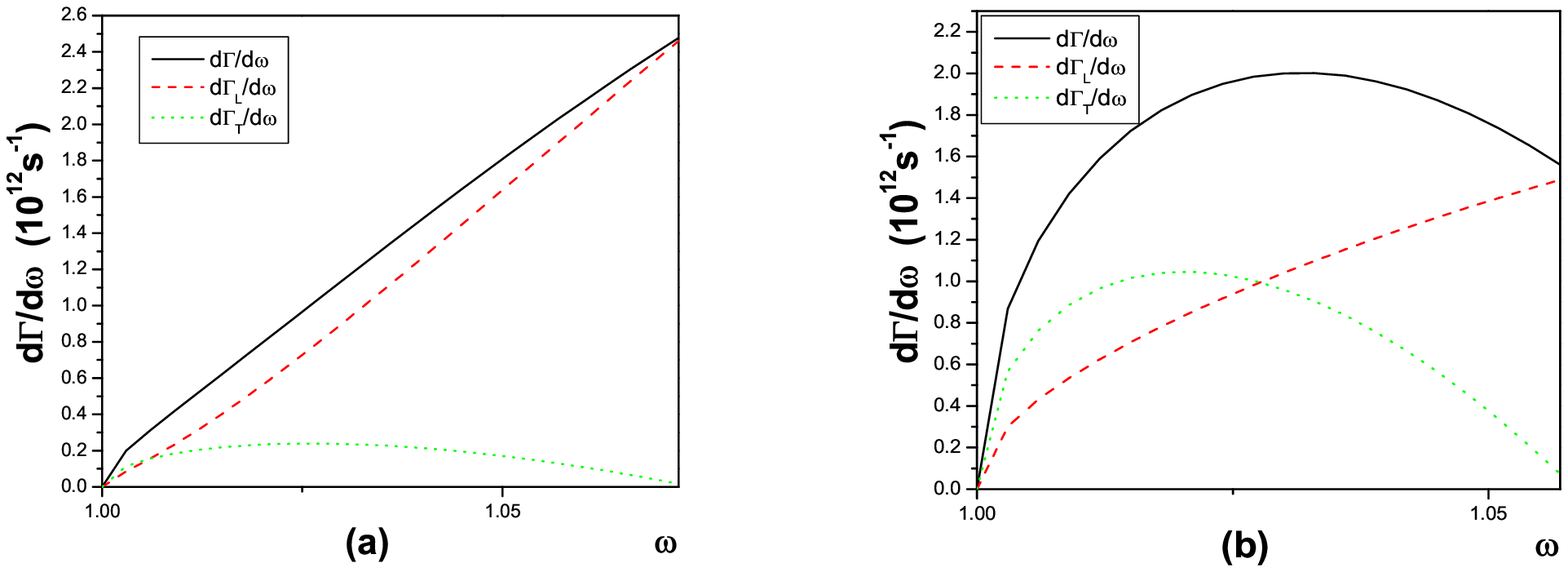}}
\end{center}
\caption{ Differential decay rates $d\Gamma/d\omega$ for the decay
$\Xi^{+}_{cc} \to \Xi^{+}_{c} l\bar{\nu}_l$(a) and $\Xi^{+}_{cc} \to \Xi^{'+}_{c}
l\bar{\nu}_l$ (b) in \cite{Ke:2021rxd} ($\theta=0^\circ$)}\label{dgtheta0}
\end{figure}

\begin{figure}
\begin{center}
%\begin{tabular}{ccc}
\scalebox{0.8}{\includegraphics{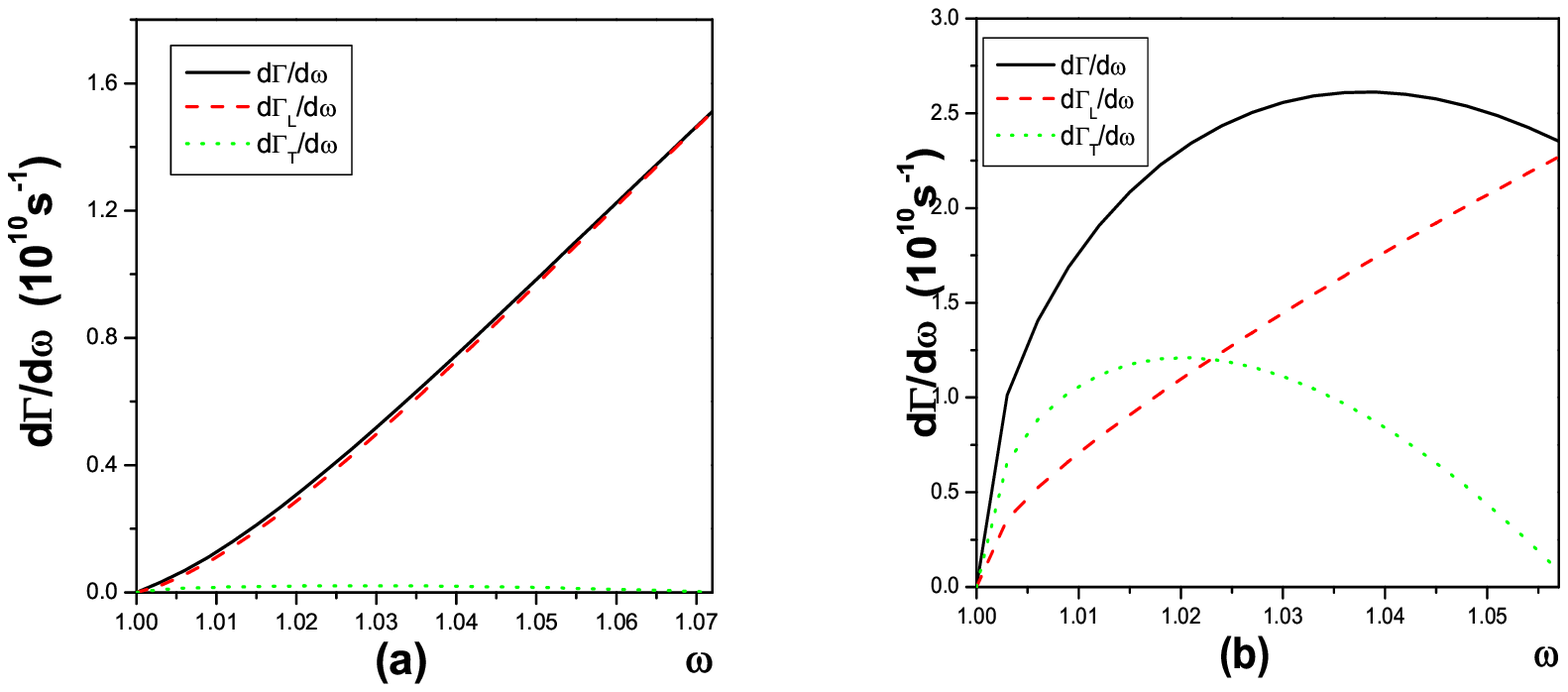}}
%\end{tabular}
\end{center}
\caption{Differential decay rates $d\Gamma/d\omega$ for the decay
$\Xi^{+}_{cc} \to \Xi^{+}_{c} l\bar{\nu}_l$(a) and $\Xi^{+}_{cc} \to \Xi^{'+}_{c}
l\bar{\nu}_l$ (b) with $\theta=16.27^\circ\pm2.30^\circ$.}\label{dgtheta1}
\end{figure}

\begin{figure}[hhh]
\begin{center}
\scalebox{0.8}{\includegraphics{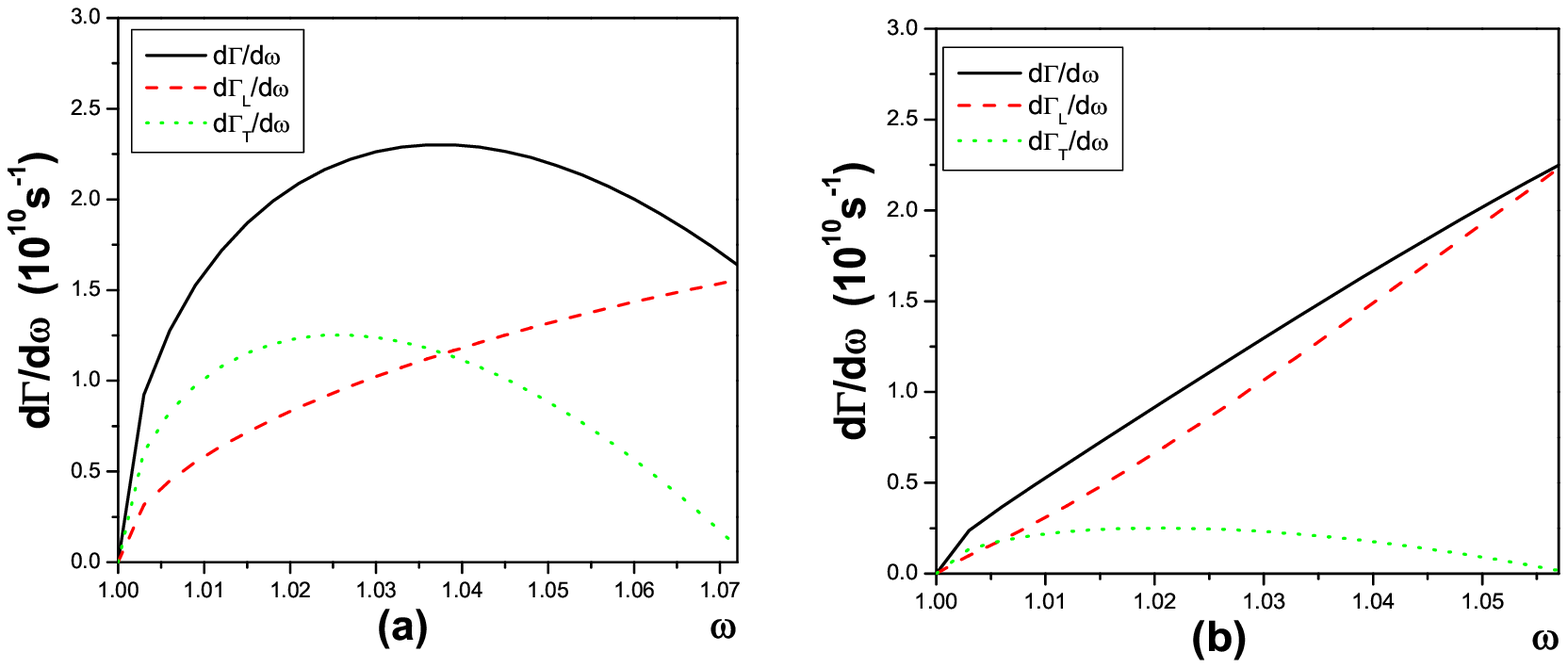}}
\end{center}
\caption{ Differential decay rates $d\Gamma/d\omega$ for the decay
$\Xi^{+}_{cc} \to \Xi^{+}_{c} l\bar{\nu}_l$(a) and $\Xi^{+}_{cc} \to \Xi^{'+}_{c}
l\bar{\nu}_l$ (b)with $\theta=85.54^\circ\pm2.30^\circ$}\label{dgtheta2}
\end{figure}

Pre-setting different mixing angles, we repeat the evaluations of
the rates of $\Xi^{+}_{cc} \to \Xi^{+}_{c} l\bar{\nu}_l$ and $\Xi^{+}_{cc} \to
\Xi^{'+}_{c} l\bar{\nu}_l$. The dependent of the differential decay widths
$d\Gamma/d\omega$ ($\omega=\frac{P\cdot P'}{mm'}$) on $\omega$ are depicted in
Fig. \ref{dgtheta0}, \ref{dgtheta1}, \ref{dgtheta2}.
One can find that the curve shapes of Fig. \ref{dgtheta0} and  Fig.\ref{dgtheta1} are similar for $\Xi^{+}_{cc} \to \Xi^{+}_{c} l\bar{\nu}_l$ and $\Xi^{+}_{cc} \to
\Xi^{'+}_{c} l\bar{\nu}_l$. By contrast, the curve tendencies  for $\Xi^{+}_{cc} \to \Xi^{+}_{c} l\bar{\nu}_l$ and $\Xi^{+}_{cc} \to
\Xi^{'+}_{c} l\bar{\nu}_l$ in Fig. \ref{dgtheta0} and in Fig. \ref{dgtheta2} are just left to right reversed from each other
while the integrated quantities (decay widths) are close.
The total decay widths and the
ratio of the longitudinal to transverse decay rates $R$ corresponding to the three mixing angles are all
listed in table \ref{Tab:t4}.

%For $\theta=0^\circ$ case the uncertainties comes from some input parameters but for $\theta=16.27^\circ\pm2.30^\circ$ and
% $\theta=85.54^\circ\pm2.30^\circ$ cases the uncertainties comes from that of $\theta$.

\begin{table}
\caption{The width (in unit $10^{12}{\rm s}^{-1}$) and the ratio $R$   of
$\Xi^{+}_{cc}\to \Xi^{+}_{c} l\bar{\nu}_l$ (left) and $\Xi^{+}_{cc}\to
\Xi^{'+}_{c} l\bar{\nu}_l$ (right).}\label{Tab:t4}
\begin{ruledtabular}
\begin{tabular}{ccc|ccc}
  &  $\Gamma$ &  R  &  $\Gamma$ & $R$    \\\hline
results in \cite{Ke:2021rxd} ($\theta=0^\circ$) &  0.100$\pm$0.015  & 7.14$\pm$0.61& 0.0995$\pm0.0091$  &  1.34$\pm$0.07\\\hline
$\theta=16.27^\circ\pm2.30^\circ$ & 0.0522$\pm$0.0051  &46.6$\pm$0.5 &0.130$\pm$0.003 &1.63$\pm$0.05\\\hline
$\theta=85.54^\circ\pm2.30^\circ$ & 0.143$\pm$0.009 &1.22$\pm0.04$
&0.0732$\pm$0.0051 &6.14$\pm$0.57
\end{tabular}
\end{ruledtabular}
\end{table}

\section{Conclusions and discussions}

In Ref.\cite{Ke:2021rxd} we calculated the transition rate of
$\Xi^{+}_{cc}\to\Xi^{(')}_{c}$ in the light front quark model with a three-quark
picture of baryon. To calculate the transition $\Xi^{+}_{cc}\to\Xi^{(')}_{c}$ we need know the inner spin-flavor structures of all the concerned baryons.
Generally the two charm quarks constitute a diquark
which joins the light quark to make the baryon $\Xi^{+}_{cc}$. Because two $c$ quarks are identical heavy flavor
fermions in a color anti-triplet, it must be a vector boson. In Ref.\cite{Ebert:2006rp} the scenario that light $us$ pair in $\Xi^{+}_{c}$ ($\Xi^{'+}_{c}$ ) is pre-set as a scalar (vector) diquark, was employed in
our previous study\cite{Ke:2021rxd}. With the prescription we calculate the form factors of the transition $\Xi^{+}_{cc}\to\Xi^{(')}_{c}$ and the decay rates of $\Xi^{+}_{cc}\to\Xi^{(')}_{c}l\bar{\nu}_l$ and
$\Xi^{+}_{cc}\to\Xi^{(')}_{c}M$.

However, we notice that the ratio $\Gamma(\Xi^{+}_{cc}\to\Xi^{+}_{c}\pi)/\Gamma(\Xi^{+}_{cc}\to\Xi^{'}_{c}\pi)$ obtained with that prescription was $0.56\pm0.18$\cite{Ke:2021rxd} which does not agree with the data newly observed by the
LHCb collaboration. To reconcile our theoretical result and data, we should find what was wrong and how to remedy the theoretical framework.
One possible pitfall might be the spin-flavor structure of $\Xi^{+}_{c}$ ($\Xi^{'+}_{c}$ ) pre-set
in Ref.\cite{Ebert:2006rp} which was based on  a precise SU(3) flavor symmetry. However, in fact the symmetry is upset by the difference between the mass of $s$ quark and those of $u$ and $d$ quarks, to manifest this breaking we are motivated to
suggest that the spin of $us$ pair in $\Xi^{+}_{c}$ ($\Xi^{'+}_{c}$ ) is the mixture of the scalar and vector diquarks. By introducing a mixing angle $\theta$ the amplitudes become
$\mathcal{A}(\Xi^{++}_{cc}\to\Xi^{+})$ = ${\rm cos}\theta\, \mathcal{A}([cc]_{1}[u]\to[us]_{0}[c])+ {\rm sin}\theta\, \mathcal{A}([cc]_{1}[u]\to[us]_{1}[c]$)
and $\mathcal{A}(\Xi^{++}_{cc}\to\Xi^{'+})$ = $-{\rm sin}\theta\, \mathcal{A}([cc]_{1}[u]\to[us]_{0}[c])+ {\rm cos}\theta\, \mathcal{A}([cc]_{1}[u]\to[us]_{1}[c]$).
%where $\mathcal{A}([cc]_{1}[u]\to[us]_{0}[c])$ and $\mathcal{A}([cc]_{1}[u]\to[us]_{1}[c]$ are the original amplitudes %$\mathcal{A}(\Xi^{++}_{cc}\to\Xi^{+})$  and
%$\mathcal{A}(\Xi^{++}_{cc}\to\Xi^{'+})$ in our early paper, respectively.
Apparently the newly achieved ratio $\Gamma(\Xi^{+}_{cc}\to\Xi^{+}_{c}\pi)/\Gamma(\Xi^{+}_{cc}\to\Xi^{'}_{c}\pi)$ depends on the
parameter $\theta$.  We fix $\theta=16.27^\circ\pm2.30^\circ$ or $85.54^\circ\pm2.30^\circ$ by fitting the data of LHCb.

Using the mixing angles we calculate the rates of semileptonic decays
$\Xi^{+}_{cc}\to\Xi^{+}_{c}l\nu_l $ and $\Xi^{+}_{cc}\to\Xi^{'}_{c}l\nu_l $.
We find that the shapes of Fig. \ref{dgtheta0} and  Fig.\ref{dgtheta1} are similar for $\Xi^{+}_{cc} \to \Xi^{+}_{c} l\bar{\nu}_l$ and $\Xi^{+}_{cc} \to
\Xi^{'+}_{c} l\bar{\nu}_l$, respectively. By contrast, the curve tendencies  for $\Xi^{+}_{cc} \to \Xi^{+}_{c} l\bar{\nu}_l$ and $\Xi^{+}_{cc} \to
\Xi^{'+}_{c} l\bar{\nu}_l$ in Fig. \ref{dgtheta0} and in Fig. \ref{dgtheta2} are just left to right reversed from each other. With the same
theoretical framework, we also evaluate the rates of several
non-leptonic decays. Our numerical results indicate that the order of magnitudes of these decays are unchanged  but
the relative sizes between $\Gamma(\Xi^{+}_{cc}\to\Xi^{+}_{c} M)$ and $\Gamma(\Xi^{+}_{cc}\to\Xi^{'}_{c} M)$ are varied for the different mixing angles.

We hope that the experimentalists can make more precise
measurements on those relevant decay channels of $\Xi^{+}_{cc}$. The new data would
tell us whether our mechanism can survive, then one can pin down the right one from the two possible mixing angles we fixed. Definitely, the
theoretical studies on the double-heavy baryons would be helpful for
getting a better understanding about the quark model and the
non-perturbative QCD effects.

\section*{Acknowledgement}

This work is supported by the National Natural Science Foundation
of China (NNSFC) under the contract No. 12075167,  11735010,
12035009 and 12075125.

\appendix

\section{Semi-leptonic decays of  $\mathcal{B}_1\to
\mathcal{B}_2  l\bar\nu_l$ }

The helicity amplitudes are related to the form factors for
$\mathcal{B}_1\to \mathcal{B}_2 l\bar\nu_l$ through the following
expressions \cite{Korner:1991ph,Bialas:1992ny,Korner:1994nh}
 \beq
 H^V_{\frac{1}{2},0}&=&\frac{\sqrt{Q_-}}{\sqrt{q^2}}\left(
  \left(\Mb+\Mc\right)f_1-\frac{q^2}{\Mb}f_2\right),\non\\
 H^V_{\frac{1}{2},1}&=&\sqrt{2Q_-}\left(-f_1+
  \frac{\Mb+\Mc}{\Mb}f_2\right),\non\\
 H^A_{\frac{1}{2},0}&=&\frac{\sqrt{Q_+}}{\sqrt{q^2}}\left(
  \left(\Mb-\Mc\right)g_1+\frac{q^2}{\Mb}g_2\right),\non\\
 H^A_{\frac{1}{2},1}&=&\sqrt{2Q_+}\left(-g_1-
  \frac{\Mb-\Mc}{\Mb}g_2\right).
 \eeq
where $Q_{\pm}=2(P\cdot P'\pm \Mb\Mc)$ and $\Mb\, (\Mc)$
represents $M_{\Xi^{+}_{cc}}$ ($M_{\Xi^{+}_{c}}$). The amplitudes for the
negative helicities are obtained in terms of the relation
 \beq
 H^{V,A}_{-\lambda'-\lambda_W}=\pm H^{V,A}_{\lambda',\lambda_W},
  \eeq
where the upper (lower) index corresponds to $V(A)$.
 The helicity
amplitudes are
 \beq
 H_{\lambda',\lambda_W}=H^V_{\lambda',\lambda_W}-
  H^A_{\lambda',\lambda_W}.
 \eeq
The helicities of the $W$-boson $\lambda_W$ can be either $0$ or
$1$, which correspond to the longitudinal and transverse
polarizations, respectively.  The longitudinally ($L$) and
transversely ($T$) polarized rates are
respectively\cite{Korner:1991ph,Bialas:1992ny,Korner:1994nh}
 \beq
 \frac{d\Gamma_L}{d\omega}&=&\frac{G_F^2|V_{cb}|^2}{(2\pi)^3}~
  \frac{q^2~p_c~\Mc}{12\Mb}\left[
  |H_{\frac{1}{2},0}|^2+|H_{-\frac{1}{2},0}|^2\right],\non\\
 \frac{d\Gamma_T}{d\omega}&=&\frac{G_F^2|V_{cb}|^2}{(2\pi)^3}~
  \frac{q^2~p_c~\Mc}{12\Mb}\left[
  |H_{\frac{1}{2},1}|^2+|H_{-\frac{1}{2},-1}|^2\right].
 \eeq
where $p_c$ is the momentum of $\mathcal{B}_2$ in the reset frame
of $\mathcal{B}_1$.

 The ratio of the longitudinal to
transverse decay rates $R$ is defined by
 \beq
 R=\frac{\Gamma_L}{\Gamma_T}=\frac{\int_1^{\omega_{\rm
     max}}d\omega~q^2~p_c\left[ |H_{\frac{1}{2},0}|^2+|H_{-\frac{1}{2},0}|^2
     \right]}{\int_1^{\omega_{\rm max}}d\omega~q^2~p_c
     \left[ |H_{\frac{1}{2},1}|^2+|H_{-\frac{1}{2},-1}|^2\right]}.
 \eeq

\section{$\mathcal{B}_1\to
\mathcal{B}_2 M$} In general, the transition amplitude of
$\mathcal{B}_1\to \mathcal{B}_2 M$ can be written as
 \beq
 {\cal M}(\mathcal{B}_1\to
\mathcal{B}_2 P)&=&\bar
  u_{\mathcal{B}_2}(A+B\gamma_5)u_{\mathcal{B}_1}, \non \\
 {\cal M}(\mathcal{B}_1\to
\mathcal{B}_2 V)&=&\bar
  u_{\mathcal{B}_2}\epsilon^{*\mu}\left[A_1\gamma_{\mu}\gamma_5+
   A_2(p_c)_{\mu}\gamma_5+B_1\gamma_{\mu}+
   B_2(p_c)_{\mu}\right]u_{\mathcal{B}_1},
 \eeq
where $\epsilon^{\mu}$ is the polarization vector of the final
vector or axial-vector mesons. Including the effective Wilson
coefficient $a_1=c_1+c_2/N_c$ (in Ref.\cite{Buras:1994ij} $a_1=1.05\pm0.10$), the decay amplitudes in the
factorization approximation are \cite{Korner:1992wi}
 \beq
 A&=&\lambda f_P(\Mb-\Mc)f_1(M^2), \non \\
 B&=&\lambda f_P(\Mb+\Mc)g_1(M^2), \non\\
 A_1&=&-\lambda f_VM\left[g_1(M^2)+g_2(M^2)\frac{\Mb-\Mc}{\Mb}\right],
 \non\\
 A_2&=&-2\lambda f_VM\frac{g_2(M^2)}{\Mb},\non\\
 B_1&=&\lambda f_VM\left[f_1(M^2)-f_2(M^2)\frac{\Mb+\Mc}{\Mb}\right],
 \non\\
 B_2&=&2\lambda f_VM\frac{f_2(M^2)}{\Mb},
 \eeq
where $\lambda=\frac{G_F}{\sqrt 2}V_{cs}V_{q_1q_2}^*a_1$ and $M$
is the meson mass. Replacing  $P$, $V$ by $S$ and $A$ in the above
expressions, one can easily obtain similar expressions for scalar
and axial-vector mesons .

The decay rates of $\mathcal{B}_1\rightarrow \mathcal{B}_2P(S)$
and up-down asymmetries are\cite{Cheng:1996cs}
 \begin{eqnarray}
 \Gamma&=&\frac{p_c}{8\pi}\left[\frac{(\Mb+\Mc)^2-M^2}{\Mb^2}|A|^2+
  \frac{(\Mb-\Mc)^2-m^2}{\Mb^2}|B|^2\right], \non\\
 \alpha&=&-\frac{2\kappa{\rm Re}(A^*B)}{|A|^2+\kappa^2|B|^2},
 \end{eqnarray}
where $p_c$ is the $\mathcal{B}_2$ momentum in the rest frame of
$\mathcal{B}_1$, $m$ is the mass of pseudoscalar (scalar),  and $\kappa=\frac{p_c}{\Mc+\sqrt{p_c^2+\Mc^2}}$. For
$\mathcal{B}_1\rightarrow \mathcal{B}_2 V(A)$ decays, the decay
rate and up-down asymmetries are
 \beq
 \Gamma&=&\frac{p_c (E_{\mathcal{B}_2}+\Mc)}{4\pi\Mb}\left[
  2\left(|S|^2+|P_2|^2\right)+\frac{\varepsilon^2}{m^2}\left(
  |S+D|^2+|P_1|^2\right)\right],\non\\
 \alpha&=&\frac{4m^2{\rm Re}(S^*P_2)+2\varepsilon^2{\rm Re}(S+D)^*P_1}
  {2m^2\left(|S|^2+|P_2|^2 \right)+\varepsilon^2\left(|S+D|^2+|P_1|^2
  \right) },
 \eeq
where $\varepsilon$ ($m$) is energy (mass) of the vector (axial
vector) meson, and
 \begin{eqnarray}
  S&=&-A_1, \non\\
  P_1&=&-\frac{p_c}{\varepsilon}\left(\frac{\Mb+\Mc}
  {E_{\mathcal{B}_2}+\Mc}B_1+M_{\mathcal{B}_1}B_2\right), \non \\
  P_2&=&\frac{p_c}{E_{\mathcal{B}_2}+\Mc}B_1,\non\\
  D&=&-\frac{p^2_c}{\varepsilon(E_{\mathcal{B}_2}+\Mc)}(A_1- M_{\mathcal{B}_1}A_2).
 \end{eqnarray}

\end{document}